\documentclass[aps,twocolumn,prb,floatfix,showpacs,superscriptaddress]{revtex4-1}
\usepackage{amsmath}
\usepackage{amssymb}
\usepackage{amsfonts}
\usepackage[pdftex]{graphicx}
\usepackage{units}
\usepackage[usenames]{color}
\usepackage{dcolumn}
\usepackage{bm}
\usepackage{float}
\usepackage[utf8]{inputenc}
\usepackage[colorlinks=true,citecolor=blue,linkcolor=blue]{hyperref}
\usepackage[normalem]{ulem}
\usepackage{xcolor}

\begin{document}

\title{Favorable band alignment for photocatalysis at the strontium germanate interface with silicon}

\author{Tom\'{a}\v{s} Rauch} 
\affiliation{Institut für Festkörpertheorie und -optik, Friedrich-Schiller-Universität Jena, Max-Wien-Platz 1, 07743 Jena, Germany}
\affiliation{European Theoretical Spectroscopy Facility}

\author{Pavel Marton}
\affiliation{Institute of Physics of the Czech Academy of Sciences, Praha, Czech Republic}
\affiliation{Institute of Mechatronics and Computer Engineering, Technical University of Liberec, Studentsk\'{a} 2, 461 17 Liberec, Czech Republic}
\author{Silvana Botti} 
\affiliation{Institut für Festkörpertheorie und -optik, Friedrich-Schiller-Universität Jena, Max-Wien-Platz 1, 07743 Jena, Germany}
\affiliation{European Theoretical Spectroscopy Facility}

\author{Ji\v{r}\'{i} Hlinka} 
\affiliation{Institute of Physics of the Czech Academy of Sciences, Praha, Czech Republic}

\date{\today}

\begin{abstract}
Photocatalytic water splitting  is a promising strategy for large-scale clean energy production. However, efficient and low-cost solid-state photocatalysts are still lacking. We present here first-principles calculations to investigate the suitability as photocathode of an epitaxial layer of strontium germanate on a Si(100) single crystal. 
Conduction and valence bands offsets at the interface between these two semiconductors were determined using state-of-the-art approximations of density functional theory for the accurate prediction of band alignments.
The resulting type-III band line-up is also confirmed by inspection of the spatially resolved density of states. 
It is concluded that the electronic structure of the investigated heterostructure is favorable for photocathodic functionality. 
\end{abstract}

\maketitle

\section{Introduction}
Solar water splitting in the presence of semiconductor photocatalysts is a clean energy source that enables renewable energy production and storage without dependence on fossil fuels and carbon dioxide emission.
Owing to its outstanding semiconductor properties and its wide use in the electronic industry, crystalline silicon remains one of the most convenient bulk materials for any optoelectronic devices.
In many applications, the native SiO$_2$ oxide layer which is purposely grown or naturally formed on the silicon surface acts as a useful mechanical protection or functional dielectric layer.

At the SiO$_2$/Si interface, the band edges of Si electronic states are contained inside the band gap of SiO$_2$.
This type of heterojunction is known as type-I \cite{FRANCIOSI19961} or straddling band lineup\cite{kroemer2001nobel}.
However, this type of interface is unsuitable for highly demanded photo-catalytic cells for hydrogen evolution reaction,  as the desired efficient and fast drift of photoexcited electrons towards the surface of the device in contact with water is hindered by the energy barrier in the conduction band and therefore another protective oxide is needed\cite{kumah2020epitaxial,spreitzer2021epitaxial,stoerzinger2018chemical,Luo2019}.
For example, it has been found that the interface between epitaxially-grown SrTiO$_3$ (STO) and a Si(100) crystalline substrate can significantly reduce the conduction-band offset\cite{PhysRevB.85.195318,Luo2019}.
In particular, high-quality epitaxial films of SrTiO$_3$ grown on a Sr-passivated Si(100) surface were identified as a direct contact of the non-dimerized (1x1) Si surface provided that no oxygen ions are contained in the interface Sr monolayer~\cite{https://doi.org/10.48550/arxiv.2202.05545}.
Their conduction band alignment is indeed favorable for the drift of the photoexcited electrons from Si to STO \cite{https://doi.org/10.48550/arxiv.2202.05545} and the catalytic performance of STO/Si based photocathodes was already demonstrated in experiments\cite{ji2014demkov}.

We explore in this Article the opportunity that similar or improved performance can be achieved with other perovskite protective oxides~\cite{doi:10.1021/acs.nanolett.2c00047,reiner2010crystalline,walter2010solar}.  In this respect, we consider the cubic perovskite form of  SrGeO$_3$ (SGO).
Recent studies showed that cubic SGO is a semiconductor with an 
indirect bandgap of about \unit[2.7]{eV} and a direct gap of about \unit[3.5]{eV}\cite{Mizoguchi2011}, a remarkably high electron mobility of the order of  \unit[400-500]{cm$^2$/Vs}\cite{Rowberg_2020} and a small effective mass at the lowest conduction band (about \unit[0.2]{$m_e$}\cite{niedermeier2020phonon}). Most interestingly, SGO belongs to the very few discovered
transparent conductive oxides \cite{Mizoguchi2011,Rowberg_2020}.
The underlying salient feature of the SGO electronic structure is that its conduction band is formed by $s$ and $p$ orbitals\cite{https://doi.org/10.1002/aelm.201800891,niedermeier2020phonon}, which are much less localized than empty $d$ orbitals of STO-like perovksites.

Throughout this paper we use the conventions of Ref.~\onlinecite{FRANCIOSI19961} to indicate band offsets at interfaces: A/B or A-B heterojunctions assume an overlayer A grown on a substrate B, and the valence-band offset (VBO) $\Delta^{\mathrm{A-B}}_{\mathrm{VBO}} $ is {\it positive} if the valence-band maximum in the overlayer (A) is {\it lower} in energy than in the substrate (B) while the conduction-band offset (CBO) $\Delta^{\mathrm{A-B}}_{\mathrm{CBO}}$ is taken {\it positive} if the conduction-band minimum in the overlayer A is {\it higher} in energy then the substrate, so that the algebraic sum of the offsets equals the bandgap difference $\Delta^{\mathrm{A-B}}_{\mathrm{VBO}}+ \Delta^{\mathrm{A-B}}_{\mathrm{CBO}} = E_{\rm g}(A)-E_{\rm g}(B)$. With such a convention, all these three energies are positive for the SiO$_2$/Si case. For applications as photocathode the CBO at the A/B interface should be zero or negative.

The central motivation for the present study is to check whether the conduction band alignment between Si and SrGeO$_3$ in SGO/Si may be favorable for releasing to water electrons that are photoexcited in the bulk silicon substrate through the protective oxide layer.
Having in mind that the character of the bonding and the electronic band structure of the two materials at the heterojunction is quite different, we tested different flavors of density functional theory.
Since the SGO/Si system has not been realized experimentally yet, and considering the similarity of SGO/Si with STO/Si we assume the STO/Si interface structure proposed in Ref.~\cite{https://doi.org/10.48550/arxiv.2202.05545}, with a passivating Sr mono-layer that separates the Si(100) unreconstructed surface from epitaxially-grown SGO.
Technical details and information on the adopted methods are given in Section~\ref{sec:comp_det}. The optimized geometry and calculated electronic properties of the interface are described in Section~\ref{sec:results}, along with the electronic structure of bulk Si and bulk SGO in an adequately strained and polarized state. 
In Section~\ref{sec:discussion} the values of the VBO and CBO are then derived independently using two different alignment procedures: the average electrostatic potential alignment\cite{PhysRevB.35.8154,PhysRevB.88.035305} (EPA) and the core level alignment~\cite{doi:10.1063/1.121249} (CLA).
These band offset calculations agree on a type-III band line-up. 
We complement them with a direct inspection of the layer-resolved projected density of states (PDOS) and local density of states (LDOS), the evaluation of charge transfer at the interface, as well as by considerations about the short-range band-bending effects. This analysis allows us to conclude at the end the suitability of the proposed epitaxial layer for application as a photocathode, as discussed in the concluding Section~\ref{sec:conclusion}.

\section{Computational details}
\label{sec:comp_det}

\subsection{\textit{Ab initio} calculations}
All DFT calculations were performed with the Vienna \textit{Ab initio} Simulation Package (VASP)~\cite{Kresse1996,KRESSE199615,PhysRevB.47.558} (version 6.2.1) that implements the projector augmented-wave method~\cite{Kresse1999}. For structural optimization we utilized the PBEsol~\cite{PhysRevLett.100.136406} exchange-correlation (XC) functional. The electronic bandstructure was then recalculated with the more accurate HSE06 hybrid functional~\cite{doi:10.1063/1.1564060,doi:10.1063/1.2204597}, for which we used the mixing value $\alpha=0.33$ to better reproduce the bulk band gap of SGO. Further explanation is provided in Sec.~\ref{sec:results:electronic_structure:bulk}. The kinetic-energy cutoff was set to \unit[500]{eV} (\unit[306.7]{eV} for bulk Si), corresponding to the \verb+PREC=High+ setting of VASP. Spin-orbit coupling was neglected in all calculations. We checked that its effect on the electronic structure of both Si and SGO is negligible for our purposes.

For supercell calculations, we used a $\Gamma$-centered grid of 12$\times$6$\times$1 \textbf{k} points for ionic relaxation (using a 2$\times$1 in-plane supercell) and 16$\times$16$\times$1 \textbf{k} points for electronic structure calculations (using a 1$\times$1 in-plane supercell). For bulk electronic structure calculations we used 12$\times$12$\times$12 and 8$\times$8$\times$8 \textbf{k} points for Si and SGO, respectively. The geometry optimization was stopped when the forces on each atom were smaller than \unit[15]{meV/\r{A}}.

In the post-processing step, we used the codes pyPROCAR~\cite{HERATH2020107080} to visualize the atom-projected band structure and density of states (PDOS) and DensityTool~\cite{LODEIRO2022108384} to obtain the local density of states (LDOS).

The figures were created with Matplotlib v2.2.2, VESTA v3.4.6~\cite{Momma:db5098}, and Inkscape v0.92.

\subsection{Interface model}
To study the atomic and electronic structure of epitxially-grown SGO on a (001) oriented Si substrate, we modeled the system as a finite slab surrounded by vacuum in a supercell. We followed the modeling procedure proposed in previous works on STO/Si heterostructures~\cite{PhysRevB.85.195318,https://doi.org/10.48550/arxiv.2202.05545}, taking advantage of the similarly to SGO/Si. For the substrate we used 9 atomic layers of Si terminated by H atoms at the interface with vacuum to simulate a semi-infinite bulk. We observed that this thickness is necessary to ensure a bulk-like behavior inside of the Si layer. The Si(100) surface was passivated by a monolayer of Sr, bonded to the SGO layer, that consisted of three unit cells terminated by an SrO layer. Based on previous results for the similar STO/Si interface, we expect the atomic and electronic structure to depend only weakly on the termination of the perovskite layer~\cite{PhysRevB.85.195318}. The in-plane lattice constant was set to the equilibrium bulk value of Si (\unit[3.843]{\r{A}})~\cite{https://doi.org/10.48550/arxiv.2202.05545}, to which the out-of-plane lattice constant of SGO has to accommodate for epitaxial growth, resulting in a substrate-induced in-plane strain of \unit[1.4]{\%}. We remark that the presence of a Sr monolayer at the STO/Si interface was proved to be the most probable reconstruction~\cite{PhysRevB.85.195318,https://doi.org/10.48550/arxiv.2202.05545}. 
Finally, the slab was embedded in more than \unit[15]{\r{A}} of vacuum.

To allow possible relaxations at the interface, we adopted a 2$\times$1 in-plane supercell during the structural relaxation of the slab (with in-plane lattice parameters $a=$ \unit[7.686]{\r{A}} and $b=$ \unit[3.843]{\r{A}}). As the final atomic structure was still symmetric, we performed the successive electronic-structure calculations with a 1$\times$1 slab model ($a=b=$ \unit[3.843]{\r{A}}).

\subsection{Band alignment calculations}
Various methods have been proposed to calculate the band alignment at the interface of two materials, ranging from pure bulk-based methods~\cite{PhysRevB.30.4874,doi:10.1063/1.3059569}, or estimate from individual ionization potentials and electron affinities that neglect the details of the common interface~\cite{Oba_2018}, to supercell calculations~\cite{PhysRevB.35.8154}. In this work, we employ the latter approach, since we want to include structural details of the interface that can have important effects on the final electronic structure. 

Since the precise numerical values of band edges at different positions in the supercell cannot be easily extracted from calculations of the electronic structure of the supercell, we perform as well two auxiliary bulk calculations (one for bulk Si and one for strained bulk SGO) and align them to chosen reference levels according to their difference across the interface in the supercell calculation.

For a sufficiently large supercell, the ionic positions are expected to reproduce the bulk crystal structure in the central part of a material layer. We assume the bond lengths in these region to build the unit cells for the auxiliary bulk calculations. Note that these auxiliary structures can be strained, and thus different from the equilibrium bulk structures due to the lattice mismatch of the two interfaced materials. As it will be further discussed in Sec.~\ref{sec:results:atomic_structure}, the Si substrate in our simulation adopts the unstrained cubic structure of Si. On the other hand, the SGO layer is very thin and it does not exhibit any periodic behavior in the direction perpendicular to the interface. We will choose an auxiliary bulk structure obtained by averaging the interlayer distances and oxygen-cation displacements in the central region of the SGO part of the supercell.

The valence band offset $ \Delta^{\mathrm{A-B}}_{\mathrm{VBO}}$ at the interface of a semiconductor A grown on a semiconducting substrate B is calculated as~\cite{PhysRevB.88.035305} 
\begin{equation}
    \Delta^{\mathrm{A-B}}_{\mathrm{VBO}} = \Delta \epsilon^{\mathrm{B}}_{\mathrm{VBM-Ref}} - \Delta \epsilon^{\mathrm{A}}_{\mathrm{VBM-Ref}} - \Delta \epsilon^{\mathrm{A-B}}_{\mathrm{Ref}},
    \label{eq:vbo}
\end{equation}
where $\Delta \epsilon_{\mathrm{VBM-Ref}}$ is the difference between the valence band maximum (VBM) and the reference level of one of the materials, obtained from a calculation of the auxiliary periodic bulk system, and $\Delta \epsilon^{\mathrm{A-B}}_{\mathrm{Ref}}$ is the difference of the reference levels in the central regions of the two semiconductors, obtained from a supercell calculation. We selected and compared two different reference levels for the band alignment at the interface: the macroscopic average of the electrostatic potential~\cite{PhysRevB.35.8154,PhysRevB.88.035305} is the chosen reference for the EPA and the average of the core level positions is the reference used for the CLA~\cite{doi:10.1063/1.121249}. 
This choice is what defines the EPA and CLA approaches mentioned in the introduction.
Both methods were previously tested in Ref.~\onlinecite{Di_Liberto_2021} and they were found in that work to give comparable results.

In the EPA, $\Delta \epsilon^{\mathrm{A-B}}_{\mathrm{Ref}} = \Delta V^{\mathrm{A-B}}$ where $V^{\mathrm{A}}$ is the macroscopic average $\overline{\overline{V}}(z)$ of the planar average $\overline{V}(z)$ of the electrostatic potential in the middle of the layer A of the supercell (denoted by the coordinate $z^{\mathrm{A}}$). The macroscopic average is calculated as
\begin{equation}
    \overline{\overline{V}}(z) = \frac{1}{\sqrt{2\pi\sigma^2}} \int \overline{V}(z') e^{-\frac{(z-z')^2}{2\sigma^2}} dz'.
    \label{eq:EPA}
\end{equation}
The Gaussian average is necessary since the interlayer distance varies for each pair of adjacent layers, in particular in the SGO part of the supercell. $V^{\mathrm{A}}$ is calculated for a range of reasonable values $\sigma \in (1.0,4.0)$\r{A}. We set as final value for further calculations the $\sigma$ for which the maximal deviation from nearby $\overline{\overline{V}}(z)$ values (for $z\in(z^{\mathrm{A}}-\sigma,z^{\mathrm{A}}+\sigma)$) is minimal. The same procedure is followed for the layer B of the supercell. 

To better understand the CLA, we rewrite Eq.~\eqref{eq:vbo} as
\begin{equation}
    \Delta^{\mathrm{A-B}}_{\mathrm{VBO}} = \epsilon^{\mathrm{B}}_{\mathrm{VBM}} - \epsilon^{\mathrm{A}}_{\mathrm{VBM}} + \Delta \epsilon^{\mathrm{B}}_{\mathrm{Ref}} - \Delta \epsilon^{\mathrm{A}}_{\mathrm{Ref}},
    \label{eq:vbo_CLA}
\end{equation}
where $\epsilon^{\mathrm{A}}_{\mathrm{VBM}}$ is the VBM of material A in bulk calculations and $\Delta \epsilon^{\mathrm{A}}_{\mathrm{Ref}}$ is the difference of the reference levels (here the core levels) of the material A in the supercell calculation and in the auxiliary bulk calculation. We calculated $\Delta \epsilon_{\mathrm{Ref}}$ for single core levels (Si:$1s$,$2s$,$2p$, Sr:$1s$,$2s$,$2p$,$3s$,$3p$,$3d$, Ge:$1s$,$2s$,$2p$,$3s$,$3p$, O:$1s$) of atoms in the central parts of the Si and SGO layers composing the supercell and we observed huge differences (up to \unit[1]{eV}) in the calculated VBOs for different choices of the reference core level. Therefore, the individual results obtained for different core levels have to be averaged appropriately. 
First, the results obtained for core levels of individual atoms are averaged. For Si the central atom in the middle of the Si layer is considered. In the case of SGO, we consider Ge and O atoms in the middle of the layer and the Sr atoms in the atomic layer next to the central one. Finally, the results for the individual atoms of SGO are averaged according to weights proportional to their number in the bulk unit cell.

Within the EPA and CLA, we can obtain the CBO from the knowledge of the VBO ($\Delta^{\mathrm{A-B}}_{\mathrm{VBO}}$) and of the band gaps of the two constituents ($E^{\mathrm{A}}_g$ and $E^{\mathrm{B}}_g$) calculated in the auxiliary bulk system:
\begin{equation}
    \Delta^{\mathrm{A-B}}_{\mathrm{CBO}} = E^{\mathrm{A}}_g - E^{\mathrm{B}}_g - \Delta^{\mathrm{A-B}}_{\mathrm{VBO}}.
    \label{eq:cbo}
\end{equation}

\section{Results}
\label{sec:results}

\subsection{Atomic structure}
\label{sec:results:atomic_structure}
\begin{figure}[ht!]
  \centering
  \includegraphics[width = 0.99\columnwidth]{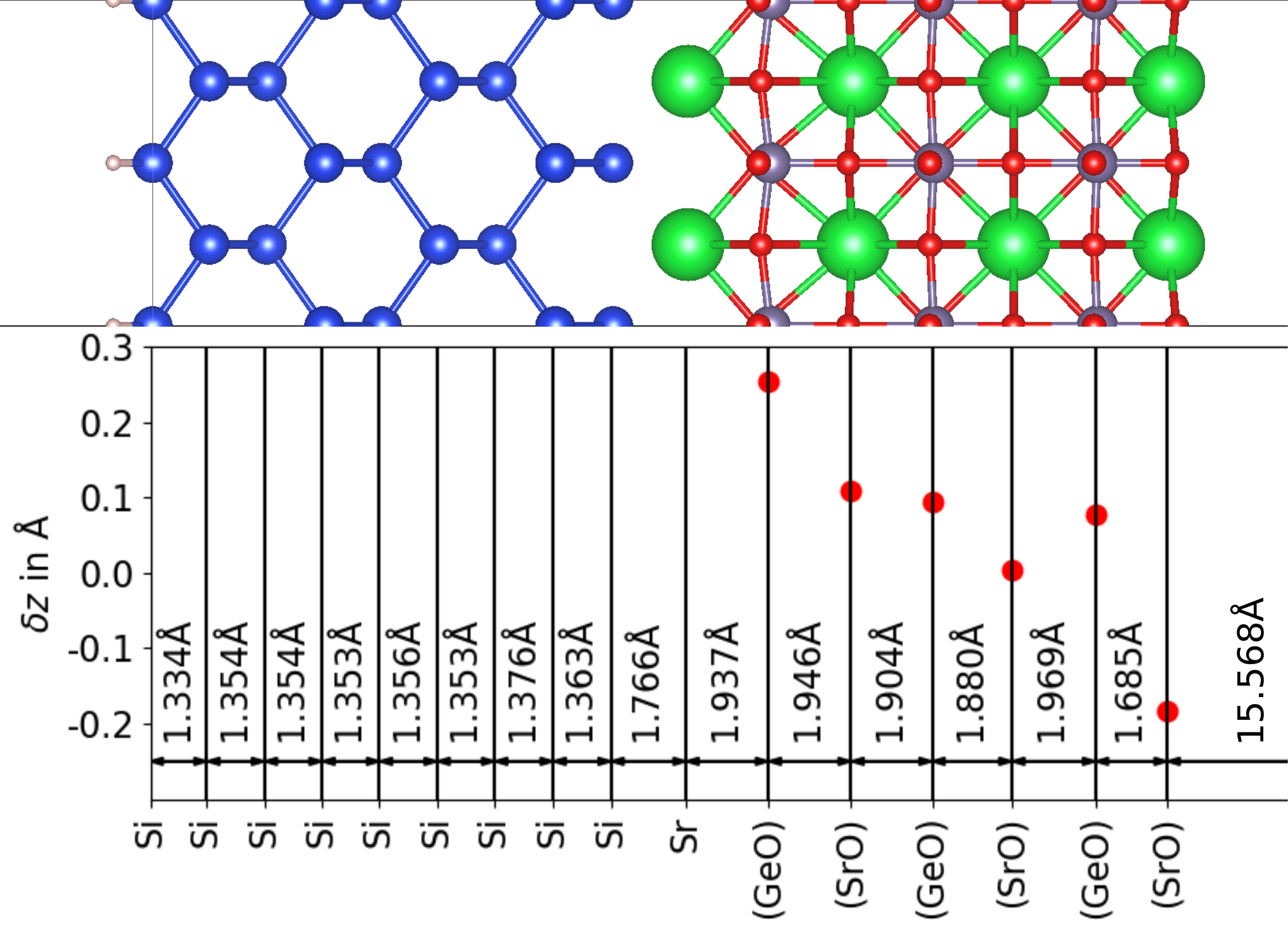}
  \caption{Relaxed atomic structure of the studied SGO/Si interface. Top: Ball-and-stick model of the structure (Si: blue, Sr: green, Ge: grey, O: red, H: beige.). Bottom: Calculated average interlayer distances (black) and the oxygen-cation distances in SGO (red dots).}
  \label{fig:figure_supercell}
\end{figure}

We display the relaxed atomic structure for the chosen SGO/Si interface in Fig.~\ref{fig:figure_supercell}. 
We can clearly observe that the average interlayer distances in the middle of the Si part of the slab are close to the Si bulk value (\unit[1.359]{\r{A}}). In the SGO part on the other hand, no bulk-like behavior can be observed. While the (GeO)-(SrO) distances are strongly influenced by the interface and surface, for the central (SrO)-(SrO) distance we obtained \unit[3.784]{\r{A}}, which indicates a strain in the SGO layer leading to a tetragonal $c/a$ ratio of 0.985. This is in agreement with the expectation based on the lattice mismatch between bulk cubic Si and cubic SGO. The average distance between the top Si layer and the first GeO layer was \unit[3.703]{\r{A}}, which is close to the result obtained previously for the same type of interface of STO/Si (\unit[3.56]{\r{A}})~\cite{PhysRevB.85.195318}.

Further, indicated by red dots in the bottom panel of Fig.~\ref{fig:figure_supercell} we show the calculated oxygen-cation displacement $\delta z$ in each atomic layer. We found $\delta z=$ \unit[0.25]{\r{A}} in the GeO-layer closest to the interface, which is again close to \unit[0.35]{\r{A}} found in the first TiO layer in STO/Si for the same type of interface~\cite{PhysRevB.85.195318}. The value of $\delta z$ decreases for deeper SGO layers and it changes sign in the surface SrO layer. Overall, we can state that the oxygen-cation displacement in a thin layer of SGO on Si substrate behaves qualitatively as in STO, but the out-of-plane strain of SGO is opposite from the one in STO/Si, since the equilibrium lattice constant of SGO (STO) is smaller (larger) than the lattice constant of Si.

Finally, to study the band alignment with the CLA or EPA, an auxiliary bulk calculation for Si and SGO is necessary. In the middle of the Si region the atomic arrangement is like the one in bulk Si and therefore the cubic equilibrium structure of Si is assumed for the auxiliary bulk calculation. Since bulk-like lattice parameters are not recovered in the thin SGO part of the supercell, we assume a strained unit cell with $c/a=0.985$ as in the center of the SGO slab. For the oxygen-cation displacement, we take the supercell value in the central GeO layer $\delta z_\mathrm{GeO}$ = \unit[0.095]{\r{A}} and the average of the two central SrO layers $\delta z_\mathrm{SrO}$ = \unit[0.056]{\r{A}}.

\subsection{Electronic structure}
\label{sec:results:electronic_structure}
\subsubsection{Bulk Si and SGO}
\label{sec:results:electronic_structure:bulk}
The electronic properties and particularly the width of the band gap are known to depend strongly on the choice of the XC functional. In the past, the hybrids PBE0~\cite{doi:10.1063/1.472933} and HSE06 with the mixing parameter $\alpha=0.33$ (called here HSE033) were employed to reproduce the experimental band gap of bulk SGO~\cite{Mizoguchi2011,C6CP05572A,Rowberg_2020}. Therefore, in Tab.~\ref{tab:band_gaps} we show the band gaps calculated with PBEsol, PBE0, HSE06, and HSE033, compared with experimental values.
\begin{table}[ht]
    \centering
    \begin{ruledtabular}
    \begin{tabular}{c|ccccc}
        & PBEsol & PBE0 & HSE06 & HSE033 & exp \\
        \hline
    Si & 0.47 & 1.78 & 1.16 & 1.36 & 1.17 \\
    SGO & 0.34 & 2.85 & 2.16 & 2.77 & 2.70 
    \end{tabular}
    \end{ruledtabular}       
    \caption{Band gaps in eV calculated with different XC functionals for bulk cubic Si ($a=$ \unit[3.843]{\r{A}}) and SGO ($a=$ \unit[3.79]{\r{A}}). The experimental values are from Ref.~\onlinecite{sze2006physics} (Si) and \onlinecite{Mizoguchi2011} (SGO).}
    \label{tab:band_gaps}
\end{table}
We observe that PBEsol severely underestimates both band gaps, as expected from a GGA XC functional. PBE0 reproduces well the experimental band gap of SGO but overestimates the Si band gap by \unit[0.61]{eV}, while HSE06 yields a very good result for Si but underestimates the SGO band gap by \unit[0.54]{eV}. Therefore, HSE033 provides the best compromise for both Si and SGO and we used it for all our electronic structure calculations in this work unless otherwise stated.

\begin{figure}[ht!]
  \centering
  \includegraphics[width = 0.99\columnwidth]{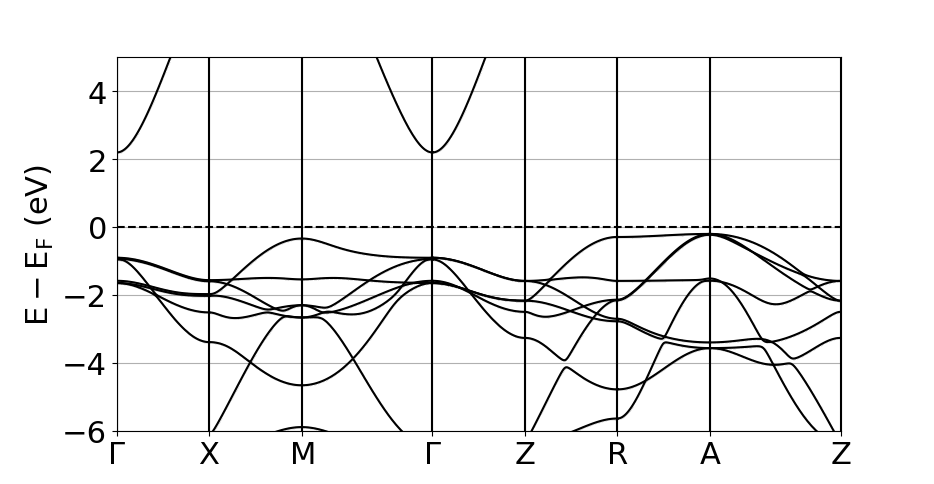}
  \caption{Electronic structure of strained SGO ($c/a=0.985$) with oxygen-cation displacements $\delta z_\mathrm{GeO}$ = \unit[0.095]{\r{A}} and $\delta z_\mathrm{SrO}$ = \unit[0.056]{\r{A}} calculated with HSE033.}
  \label{fig:figure_bands_SGO_HSE}
\end{figure}

For the evaluation of the band alignment in the SGO/Si heterostructure with EPA and CLA we used the strained and polarized crystalline auxiliary SGO unit cell, as described in Sec.~\ref{sec:results:atomic_structure}. For this structure, we obtained for the band gap \unit[0.06]{eV} with PBEsol and \unit[2.4]{eV} with HSE033. The band gap thus becomes smaller with strain. We show the electronic band structure of the strained and polarized bulk SGO calculated with HSE033 in Fig.~\ref{fig:figure_bands_SGO_HSE}.
The band gap is indirect and the lowest conduction band at $\Gamma$ has a small effective mass and DOS, as was found earlier in Ref.~\onlinecite{Rowberg_2020}. The highest valence bands, on the other hand have a much larger DOS.

\subsubsection{SGO/Si interface}
In this section we finally present the main result of this work - the calculated electronic structure of a thin SGO layer on a Si substrate with the chosen model interface including a full Sr monolayer. We recall that in our convention the VBO is positive when the VBM of SGO lies below the one of Si and the CBO is negative when the conduction band minimum (CBM) of SGO lies below the one of Si.

The calculations of the EPA and CLA were obtained by means of Eqs.~\eqref{eq:vbo}, \eqref{eq:EPA}, \eqref{eq:vbo_CLA} and \eqref{eq:cbo}, using the auxiliary bulk band edge positions. We summarize our final VBOs and CBOs, calculated with the PBEsol and HSE033 XC functionals, in the first two columns of Tab.~\ref{tab:vbos}.
\begin{table}[ht]
    \centering
    \begin{ruledtabular}
    \begin{tabular}{c|ccc}
    \multicolumn{4}{c}{Valence band offsets}\\
    \hline
        & EPA & CLA & PDOS  \\
        \hline
    PBEsol & 2.09 & 2.18 & 2.12  \\
    HSE033 & 3.31 & 3.44 & 3.76  \\
    \hline
        \multicolumn{4}{c}{Conduction band offsets}\\
    \hline
        & EPA & CLA & PDOS  \\
        \hline
    PBEsol & -2.51 & -2.60 & -1.32 \\
    HSE033 & -2.26 & -2.39 & -1.89 \\
    \hline
    \multicolumn{4}{c}{$\epsilon^{\mathrm{SGO}}_{\mathrm{CBM}} - \epsilon^{\mathrm{Si}}_{\mathrm{VBM}}$}\\
    \hline
        & EPA & CLA & PDOS  \\
        \hline
    PBEsol & -2.03 & -2.13 & -0.63  \\
    HSE033 & -0.91 & -1.04 & -0.43  \\
    \end{tabular}
    \end{ruledtabular}       
    \caption{Valence band offset ( $\Delta_{\mathrm{VBO}}$), conduction band offset ($\Delta_{\mathrm{CBO}}$), and the distance between the CBM of SGO and VBM of Si obtained with the EPA, CLA, and estimated from the PDOS (Fig.~\ref{fig:figure_pdos_2Si_2SGO}) in eV.}
    \label{tab:vbos}
\end{table}
In addition to the VBO and CBO, we also provide in Tab.~\ref{tab:vbos} the calculated difference of the CBM of SGO and VBM of Si.

\section{Discussion}
\label{sec:discussion}

\subsection{Band alignment}
We begin the discussion of our results with the calculated VBOs. The results obtained with EPA and CLA using the same XC functional are in excellent agreement. The average $\Delta_{\mathrm{VBO}}=$ \unit[3.38]{eV} from HSE033 is significantly larger than the average $\Delta_{\mathrm{VBO}}=$ \unit[2.14]{eV} from PBEsol, showing that the more precise hybrid XC functional is necessary for this calculation. 

Also in the case of CBO, the EPA and CLA results are very similar, with an average value of $\Delta_{\mathrm{CBO}}=$ \unit[-2.33]{eV} (\unit[-2.56]{eV}) with HSE033 (PBEsol). Interestingly, we note that the PBEsol and HSE033 functionals yield similar results for the conduction band offsets, even though the calculated band gaps obtained with the different XC functionals differ considerably. We will discuss in Sec.~\ref{sec:ldos} that the CBO is almost independent of the XC functional because the CBM of SGO is pinned to the Fermi level of the supercell. A similar effect was found previously also for interface models of STO/Si~\cite{PhysRevB.85.195318}.

Comparing the calculated CBOs with the Si band gap, we observe that the CBM of SGO ($\epsilon^{\mathrm{SGO}}_{\mathrm{CBM}}$) lies below the VBM of Si ($\epsilon^{\mathrm{Si}}_{\mathrm{VBM}}$), indicating an overall metallic system. To obtain a measure for how far the system is from being an insulator, we also present the values of $\epsilon^{\mathrm{SGO}}_{\mathrm{CBM}} - \epsilon^{\mathrm{Si}}_{\mathrm{VBM}}$ in Tab.~\ref{tab:vbos}. 
Also here, EPA and CLA yield very close results, but the choice of the XC functional changes the result by $\sim$\unit[1]{eV}. On average, the CBM of SGO is calculated to be located \unit[0.98]{eV} (\unit[2.08]{eV}) below the VBM of Si with HSE033 (PBSEsol).

\subsection{Local density of states}
\label{sec:ldos}
\begin{figure}[ht!]
  \centering
  \includegraphics[width = 0.99\columnwidth]{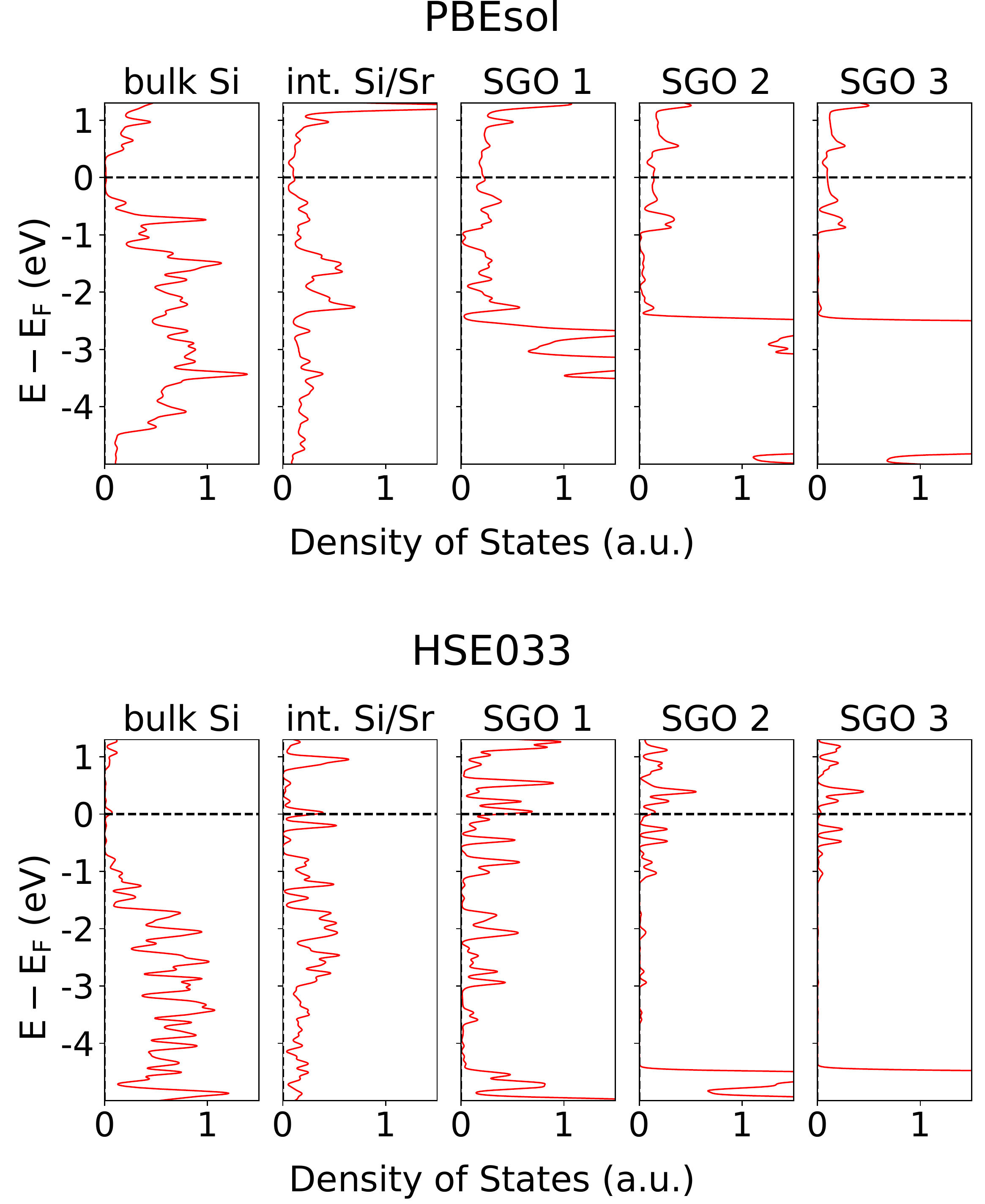}
  \caption{PDOS of chosen atomic layers of the SGO/Si supercell calculated with PBEsol (top) and HSE033 (bottom).}
  \label{fig:figure_pdos_2Si_2SGO}
\end{figure}

\begin{figure}[ht!]
  \centering
  \includegraphics[width = 0.99\columnwidth]{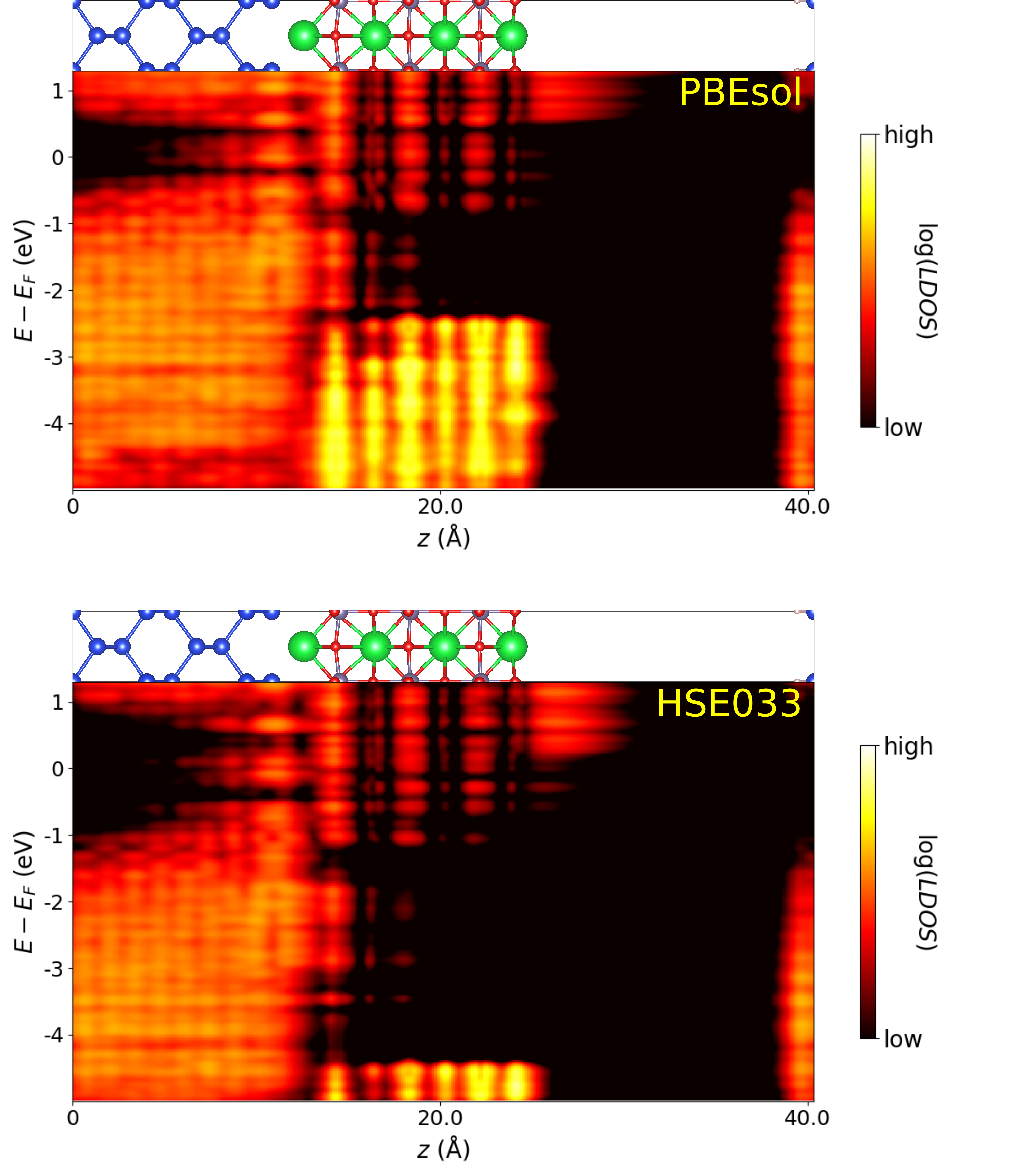}
  \caption{Atomic structure and LDOS of the SGO/Si supercell calculated with PBEsol (top) and HSE033 (bottom).}
  \label{fig:figure_ldos_2Si_2SGO}
\end{figure}

In order to visualize and interpret the local band structure of the heterostructure near the interface, we calculated the spatially resolved density of states in the form of PDOS for atoms belonging to chosen layers of the supercell (shown in Fig.~\ref{fig:figure_pdos_2Si_2SGO}) and the in-plane integrated LDOS~\cite{LODEIRO2022108384} (shown in Fig.~\ref{fig:figure_ldos_2Si_2SGO}): 
\begin{equation}
    \bar{L}(E,z) = \frac{\Omega_{\mathrm{cell}}}{(2\pi)^3} \sum_n \int_{\mathrm{BZ}} \delta(E-\epsilon_{n,\mathbf{k}})\bar{P}_{n,\mathbf{k}}(z)\ d^{3}k
    \label{eq:ldos}
\end{equation}
where $\Omega_{\mathrm{cell}}$ is the unit cell volume, $E$ the energy and $\bar{P}_{n,\mathbf{k}}(z)$ the in-plane integrated partial charge density $P_{n,\mathbf{k}}(\mathrm{r})=\left|\varphi_{n,\mathbf{k}}(\mathrm{r})\right|^2$ with the Kohn-Sham wave functions $\varphi_{n,\mathbf{k}}(\mathrm{r})$ and eigenvalues $\epsilon_{n,\mathbf{k}}$. 

PDOS and LDOS are usually not suitable for a precise extraction of the band offsets, since (i) states originating from one constituent, the interface, or the surface often reach into the band gap of the second constituent, making the identification of band edges complicated, (ii) the size of the local band gaps in the slab is larger than in the auxiliary bulk calculation due to the finite size of the supercell, and (iii) the DOS is broadened due to the finite $\mathbf{k}$-point sampling for sums in the reciprocal space and due to the practical substitution of the delta-function in Eq.~\eqref{eq:ldos} by a Gaussian. 

Therefore, it is better to resort to EPA or CLA using the electronic structure from auxiliary bulk calculations to obtain accurate band alignments for large heterostructures. On the other hand, if one is interested in the local electronic states close to the interface or within thin layers, a direct inspection of PDOS or LDOS can be more instructive.
We estimated the band-edge positions from the PDOS of the supercell, knowing that the finite smearing can introduce an error of $\sim$\unit[0.1-0.2]{eV} due to the choice of the smearing parameter $\sigma=$ \unit[0.05]{eV} both in the self-consistent calculation as well as in the LDOS calculation with Eq.~\eqref{eq:ldos}.
We used the PDOS of the three central Si layers (``bulk Si'' in Fig.~\ref{fig:figure_pdos_2Si_2SGO}) and the central SGO layer (``SGO2'' in Fig.~\ref{fig:figure_pdos_2Si_2SGO}) as representative of the two materials forming the interface. The extracted local band gaps of Si (PBEsol: $E_g^{\mathrm{Si,slab}}=$ \unit[0.69]{eV}, HSE033: $E_g^{\mathrm{Si,slab}}=$ \unit[1.46]{eV}) are fairly close to the bulk value.
On the other hand, for SGO we obtained $E_g^{\mathrm{SGO,slab}}=$ \unit[1.49]{eV} with PBEsol and $E_g^{\mathrm{SGO,slab}}=$ \unit[3.33]{eV} with HSE033. These values are 
$\sim$\unit[1.0-1.5]{eV}
larger than the band gap of the auxiliary bulk SGO structure used in the EPA and CLA. 
A band gap broadening at the order of 1\,eV is indeed expected already from the simple model of a particle with an effective mass in an infinite box potential with a width corresponding to the width of the thin SGO layer~\cite{Fox2007}.
This shows that the effect of quantum confinement is very strong for the thin SGO layer in our supercell and the auxiliary bulk structure is not suitable to calculate the positions of the SGO band edges in the thin SGO layer.
The VBO, CBO, and $\epsilon^{\mathrm{SGO}}_{\mathrm{CBM}} - \epsilon^{\mathrm{Si}}_{\mathrm{VBM}}$ obtained from the PDOS are listed in the last column of Tab.~\ref{tab:vbos}. We observe a large discrepancy between the PDOS-derived values and those obtained with EPA and CLA in case of CBO and $\epsilon^{\mathrm{SGO}}_{\mathrm{CBM}} - \epsilon^{\mathrm{Si}}_{\mathrm{VBM}}$, whereas all three methods agree for the VBO.

\begin{figure}[ht!]
  \centering
  \includegraphics[width = 0.99\columnwidth]{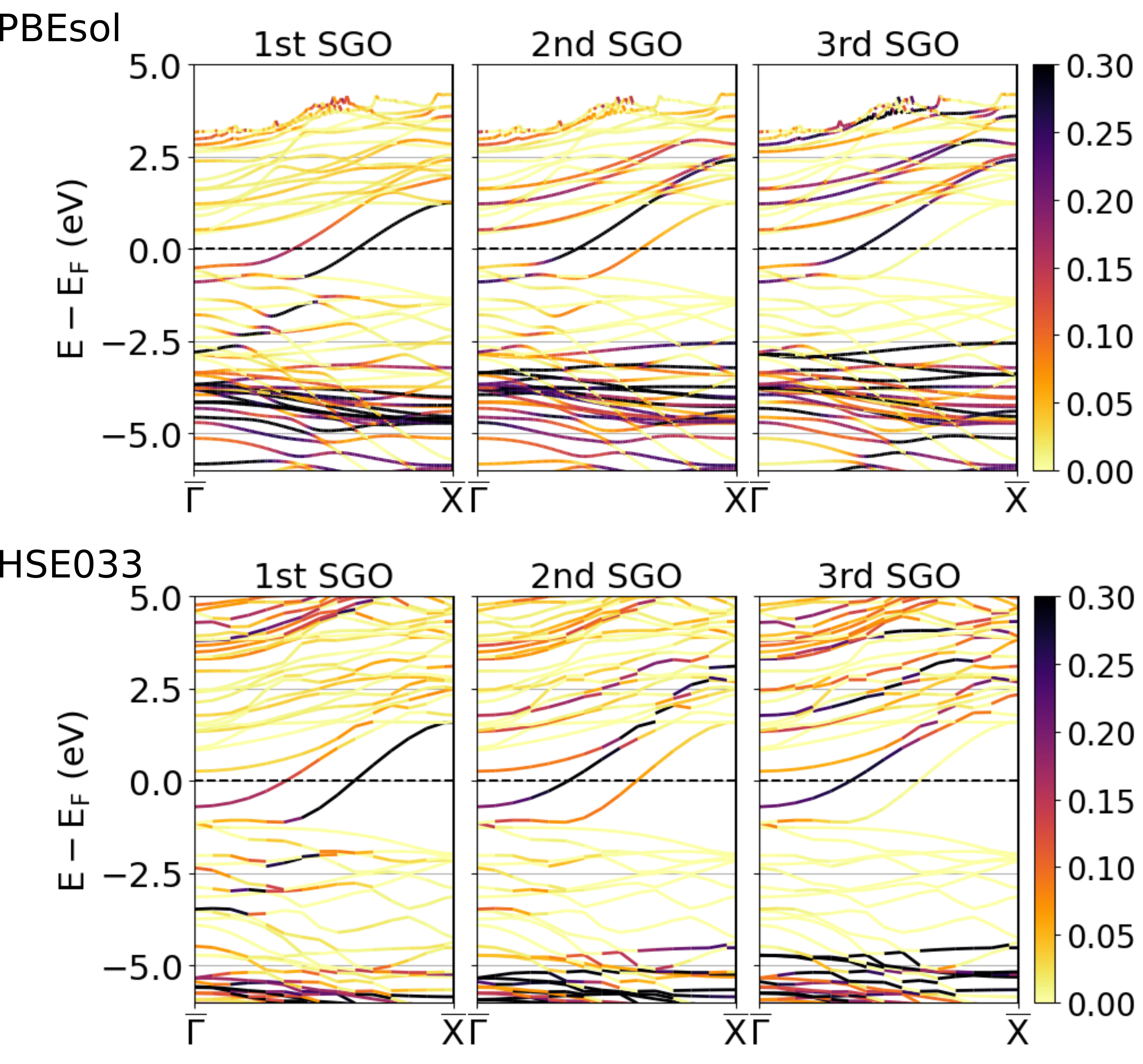}
  \caption{Band structure of the SGO/Si supercell calculated with PBEsol (top) and HSE033 (bottom). The color code shows the projection of the calculated wave functions for bands close to the gap on atoms of chosen SGO layers in the supercell.}
  \label{fig:figure_bands_projected}
\end{figure}

To gain more insight into the electronic structure of the supercell, we present in Fig.~\ref{fig:figure_ldos_2Si_2SGO} the calculated LDOS, that we plot together with the structural model. While the general information is similar to the one offered by the PDOS, the LDOS provides a much better spatial resolution. Most importantly, we can see the metallic character of the interface, which is caused by the fact that the lowest conduction band of SGO is located below $E_F$ and thus partially occupied. This has to be counterbalanced by unoccupied valence bands of Si, which is also visible in the LDOS in the Si side of the interface. This band bending is particularly interesting in case of SGO. In Fig.~\ref{fig:figure_bands_projected} we show the band structure of the whole supercell with a color code indicating the projection of the Kohn-Sham wave functions onto different SGO layers in the supercell.
The conduction band of SGO with the large band velocity (visible along $\Gamma$-X also in the bulk band structure in Fig.~\ref{fig:figure_bands_SGO_HSE}) is located fairly low in the first SGO layer, with its minimum in the vicinity of the local VBM with PBEsol and $\sim$\unit[1.5]{eV} above it with HSE033. In the second and third SGO layer, this SGO conduction band is located much higher in energy (close to $E_F$) with both XC functionals.
On the contrary, the VBM of SGO remains almost at the same energy in all three SGO layers. It is thus only the CBM of SGO which exhibits band bending.

\subsection{Interface charge transfer}

As described above, the formation of the studied interface between the Si substrate and the thin layer of SGO leads to the depletion of the Si valence bands and occupation of SGO conduction bands. 
We calculated the magnitude of the transferred charge and the generated interface dipole from the \textit{ab-initio} induced charge density
\begin{equation}
    \delta\rho(x,y,z) = \rho_{\mathrm{tot}}(x,y,z) - \rho_{\mathrm{Si/Sr}}(x,y,z) - \rho_{\mathrm{SGO}}(x,y,z) \,,
    \label{eq:charge_transfer}
\end{equation}
where $\rho_{\mathrm{tot}}$ is the charge density of the full supercell model, $\rho_{\mathrm{Si/Sr}}$ and $\rho_{\mathrm{SGO}}$ are the charge densities of the separated Si substrate (including the interface Sr monolayer) and the thin SGO layer, respectively. We used the same (arbitrary) choice for the plane dividing the two constituents as in Ref.~\onlinecite{PhysRevB.85.195318}. We further integrate the induced charge density in the plane of the interface and obtain $\Delta \rho (z) = \int \delta\rho(x,y,z) d\!x d\!y$. Finally, we calculate the induced charge $\Delta Q = \int^{\mathrm{vac. SGO}}_{\mathrm{int}} \Delta \rho (z) dz$ and interface dipole $p_{\mathrm{int}} = \int^{\mathrm{vac. SGO}}_{\mathrm{vac. Si}} \Delta \rho (z) z dz$ where ``int'' indicates the origin ($z=0$) which is set to be the plane at the interface for which $\Delta \rho$ crosses zero~\cite{https://doi.org/10.1002/aelm.201800891,doi:10.1021/acs.jctc.1c00255} and ``vac. Si'' and ``vac. SGO'' are positions in the vacuum region on the Si and SGO side of the slab, respectively, where $\Delta \rho = 0$. The calculation of the transferred charge and interface dipole was done with PBEsol. Finally, for the studied interface of SGO/Si, we obtained \mbox{$\Delta Q =$ \unit[0.14]{e}} and \mbox{$p_{\mathrm{int}}=$ \unit[0.19]{e\r{A}}} per $1\times 1$ interface unit cell (\unit[3.843]{\r{A}} $\times$ \unit[3.843]{\r{A}}). The calculated interface dipole of SGO/Si is considerably smaller than the calculated interface dipole of the same interface model of STO/Si~\cite{PhysRevB.85.195318} (\unit[0.69]{e\r{A}} per $1\times 1$ interface unit cell), but it is in the general range of values common in STO/Si interfaces.

\subsection{Photocatalytic properties}
\begin{figure}[ht!]
  \centering
  \includegraphics[width = 0.99\columnwidth]{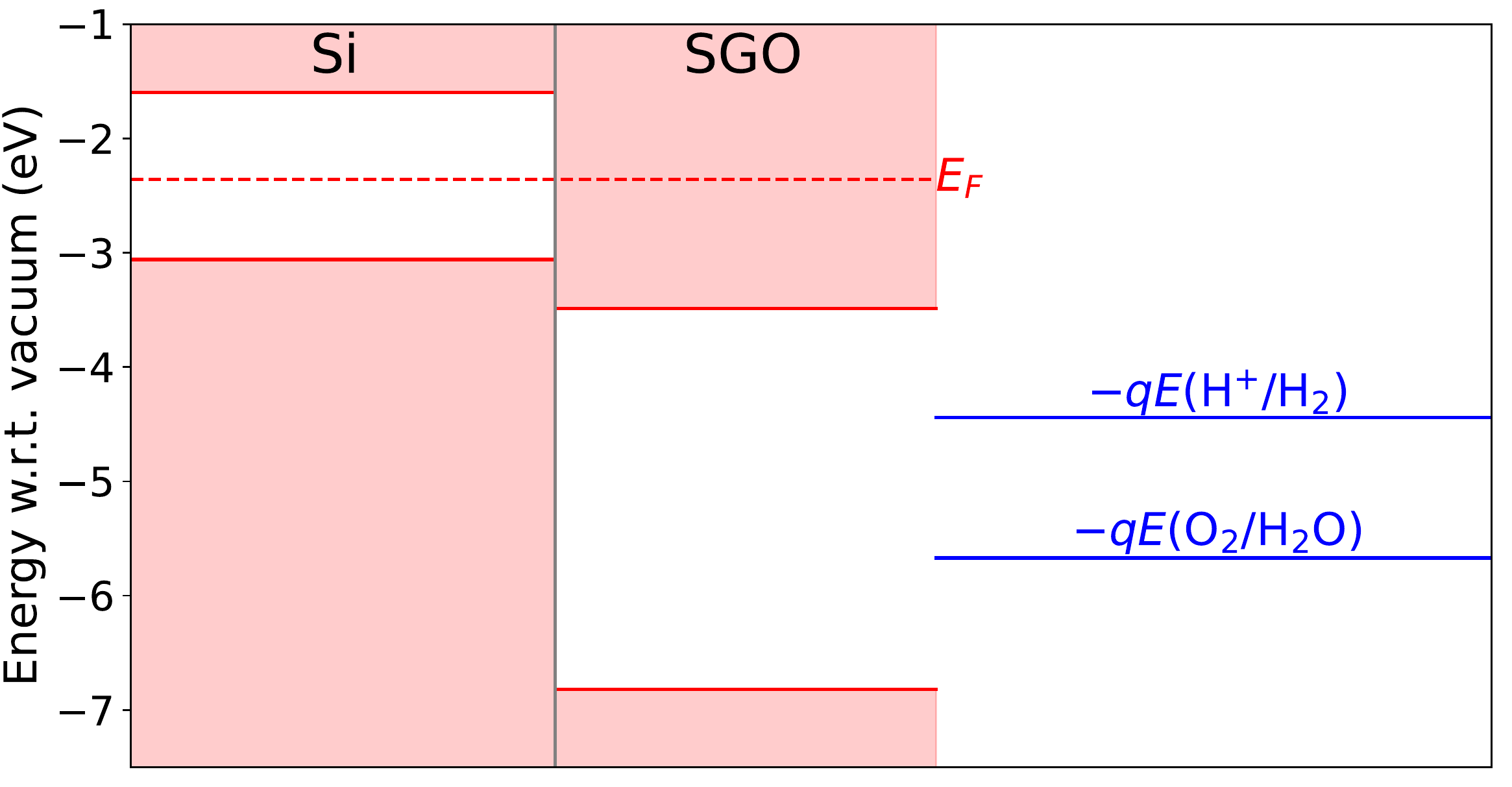}
  \caption{SGO/Si band edges aligned against vacuum potential together with the redox potentials of water splitting reactions. The band edges from the PDOS HSE033 result are shown.}
  \label{fig:figure_redox}
\end{figure}

After the successful demonstration of STO/Si as a photocathode for water reduction in Ref.~\onlinecite{ji2014demkov}, we want to consider the SGO/Si platform for the same application. We aligned therefore the band structure of the supercell and the redox potentials of H$^{+}$/H$_2$ and O$_2$/H$_2$O against the vacuum level. The PDOS result from the HSE033 calculation is schematically represented in Fig.~\ref{fig:figure_redox} as we believe it is the most accurate for the thin SGO layer on Si substrate.
We observe that both redox potentials lie inside the local band gap of SGO and below the Fermi energy of the supercell. Therefore, photogenerated electrons from the bulk of Si coming to the SGO-capped surface are expected to pass the SGO/Si interface and participate in water reduction. We conclude that the thin SGO layer with partially occupied lowest conduction band would play the role of a metallic electrode in the system.

\section{Conclusions}
\label{sec:conclusion}
We studied the structural and electronic properties of a thin SGO layer on a (100)-terminated Si substrate. We found the oxygen-cation displacement in the atomic layers of SGO qualitatively similar to the same interface structure of STO/Si~\cite{PhysRevB.85.195318}.
We calculated the band alignment using either a combination of a supercell calculation with auxiliary bulk calculations (EPA, CLA) or directly from the electronic structure of the supercell (PDOS, LDOS).
All three methods agree in the identification of a type III~\cite{FRANCIOSI19961} band alignment and the whole system is thus metallic. 
A similar electronic structure with type III band alignment has been recently measured experimentally for a STO/Si interface and reported in Ref.~\onlinecite{PhysRevMaterials.5.104603}.

Due to the strong broadening effect of the quantum confinement on the local band gap of the thin SGO layer, the PDOS is more appropriate to study the quantitative band alignment of the studied system. With the custom HSE033 hybrid functional, we found a large VBO of \unit[3.76]{eV} and CBM of SGO being located \unit[0.43]{eV} below the VBM of Si. We further found a charger transfer from Si to SGO. This causes the VBM of Si to be bent up above $E_F$ and depopulated at the interface, whereas the CBM of SGO is bent down and populated at the interface, as can be seen in PDOS (Fig.~\ref{fig:figure_pdos_2Si_2SGO}), LDOS (Fig.~\ref{fig:figure_ldos_2Si_2SGO}), and the projected band structure (Fig.~\ref{fig:figure_bands_projected}).

Finally, the local band gap of the thin SGO layer contains both redox potentials of water splitting, and thus photogenerated electrons from Si bulk could pass the interface and participate in hydrogen production at the SGO surface.

Our results demonstrate that SGO might be an alternative to STO as a capping layer on Si with a CBO favorable for the transition of excited electrons from Si to SGO. In order to further understand the structural and electronic properties of real SGO/Si interfaces, experimental evidence for the interface atomic structure, as well as further theoretical calculations are necessary. Nevertheless, our work provides first hints that the properties of the SGO/Si are similar to STO/Si and therefore this system has a great potential for applications as photocathode.

\begin{acknowledgments}
This work was supported by the Czech Science Foundation (project no. 21-20110K).
T.R. and S.B. acknowledge funding from the Volkswagen Stiftung (Momentum) through the project ``dandelion''. 
J.H., T.R. and P.M. thank Dr. Wen-Yi Tong and prof. Philippe Ghosez for their kind advice and sharing POSCAR files from their recent
work~\cite{https://doi.org/10.48550/arxiv.2202.05545} prior its publication.
Computational resources were supplied by the project ``e-Infrastruktura CZ'' (e-INFRA CZ LM2018140 ) supported by the Ministry of Education, Youth and Sports of the Czech Republic. 

\end{acknowledgments}


\bibliographystyle{apsrev4-1}
\bibliography{main.bbl}


\end{document}